\documentclass{article}

\usepackage{amsmath, amssymb, graphics, graphicx}
\usepackage{textcomp, subfigure}

\usepackage{color, soul}
\usepackage[strict]{changepage}
\usepackage{tikz}
\usepackage{cite}

\usepackage{ragged2e}

\usepackage[labeled]{multibib}
\newcites{S}{Supplementary References}

\usepackage{titling}

\begin{document}

\title{Observation of Spontaneous Brillouin Cooling}

\author{Gaurav Bahl$^{1\ast}$, Matthew Tomes$^1$, Florian Marquardt$^{2,3}$, Tal Carmon$^1$\\
\\
\footnotesize{$^1$Electrical Engineering and Computer Science, University of Michigan,}\\
\footnotesize{Ann Arbor, Michigan, USA}\\
\footnotesize{$^2$Institut f\"ur Theoretische Physik, Universit\"at Erlangen-N\"urnberg,}\\
\footnotesize{Staudtstrasse 7, D-91058 Erlangen, Germany}\\
\footnotesize{$^3$Max Planck Institute for the Science of Light,}\\
\footnotesize{G\"unther-Scharowsky-Strasse 1/Bau 24, D-91058 Erlangen, Germany}\\
%\\
\footnotesize{$^\ast$To whom correspondence should be addressed; E-mail: bahlg@umich.edu.}
}
\date{}
\maketitle

\begin{abstract}
While radiation-pressure cooling is well known \cite{Arcizet:2006p1092, Gigan:2006p1091, Kleckner:2006p1082, Riviere:2011cj, Chan_GroundState_2011}, the Brillouin scattering of light from sound is considered an acousto-optical amplification-only process \cite{PhysRevLett.12.592,Boyd,Shelby:1985p1169,GrudininCaF2lasing,Tomes2009,Bahl:2011cf}. It was suggested that cooling could be possible in multi-resonance Brillouin systems \cite{Grudinin:2010fe,GrudininCaF2lasing,Tomes2009,Bahl:2011cf} when phonons experience lower damping than light \cite{Grudinin:2010fe}. However, this regime was not accessible in traditional Brillouin systems \cite{PhysRevLett.12.592,Boyd,Shelby:1985p1169,GrudininCaF2lasing,Tomes2009} since backscattering enforces high acoustical frequencies associated with high mechanical damping \cite{Boyd}. Recently, forward Brillouin scattering \cite{Shelby:1985p1169} in microcavities \cite{Bahl:2011cf} has allowed access to low-frequency acoustical modes where mechanical dissipation is lower than optical dissipation, in accordance with the requirements for cooling \cite{Grudinin:2010fe}. Here we experimentally demonstrate cooling via such a forward Brillouin process in a microresonator. We show two regimes of operation for the Brillouin process: acoustical amplification as is traditional, but also for the first time, a Brillouin cooling regime. Cooling is mediated by an optical pump, and scattered light, that beat and electrostrictively attenuate the Brownian motion of the mechanical mode.
\end{abstract}

\begin{figure}[p]  %[h!]
	\begin{adjustwidth}{-1in}{-1in}
	 \centering
		\subfigure[] { 
			\includegraphics[width=1.2\hsize, clip=true, trim=0in 5in 0in 0in]{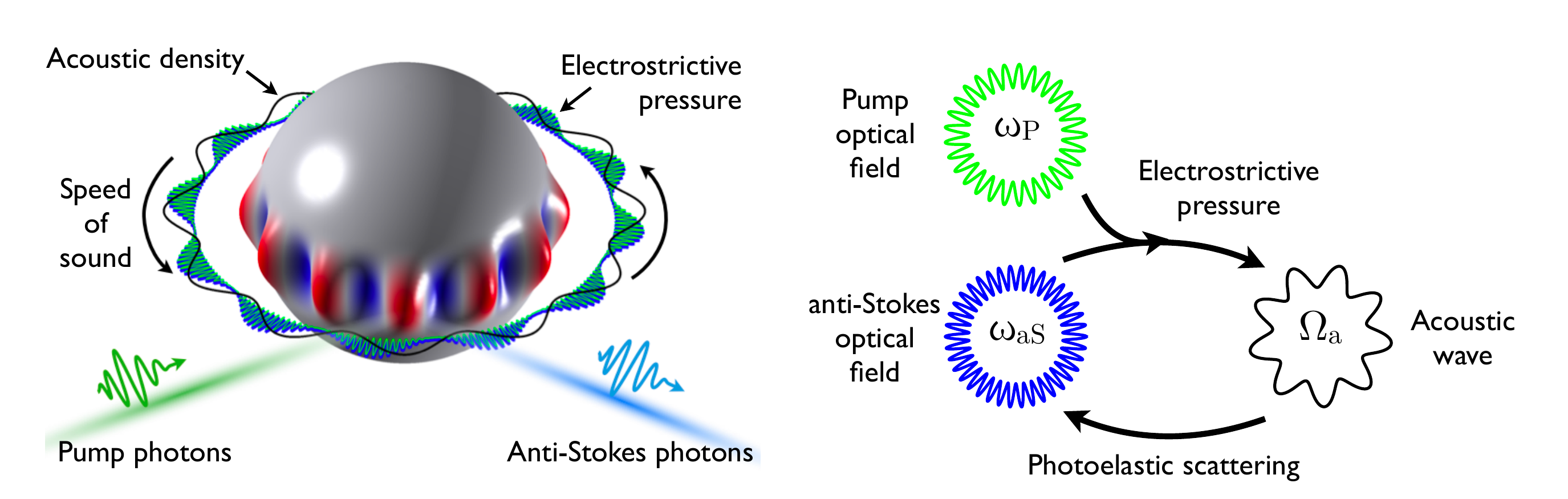}  %0.55   4.35		
		}
		\subfigure[] {
			\includegraphics[width=0.65\hsize, clip=true, trim=0.3in 2.6in 0.5in 1.1in]{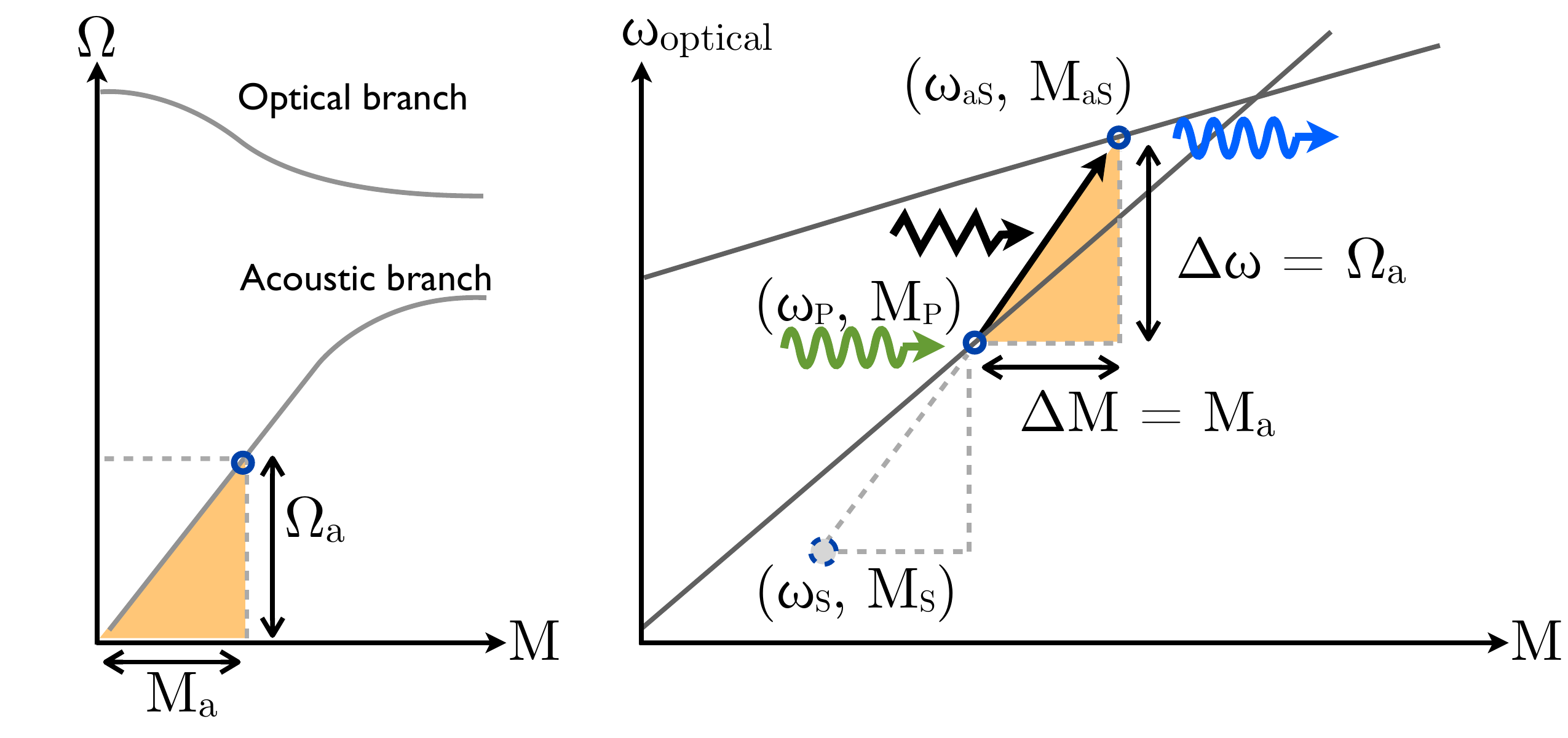}  %0.68
		}  %0.65
		\subfigure[] {
			\includegraphics[width=0.72\hsize, clip=true, trim=0.2in 1.6in 0.5in 1.4in]{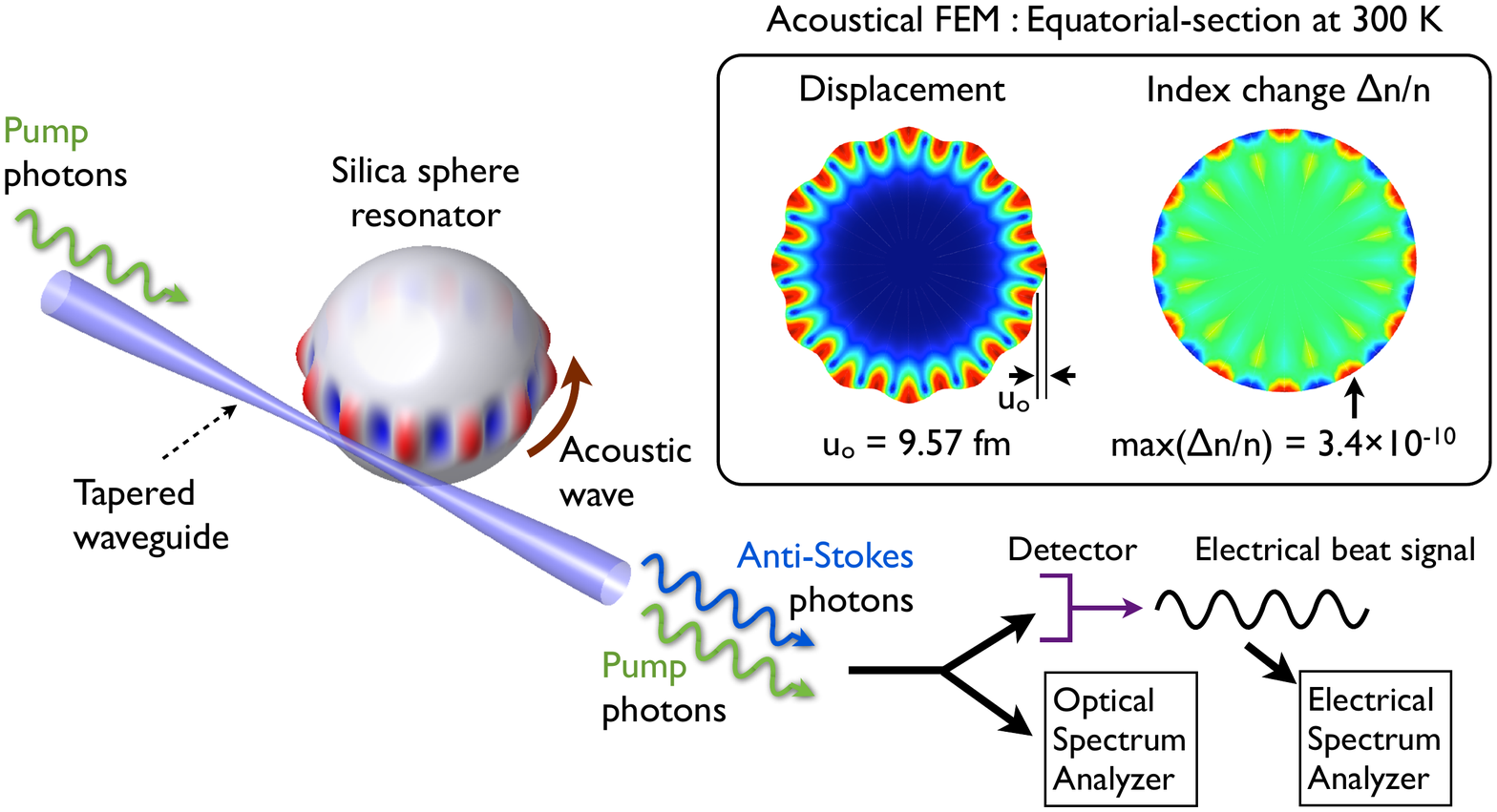}   
		}
		\subfigure[] {
			\includegraphics[width=0.7\hsize]{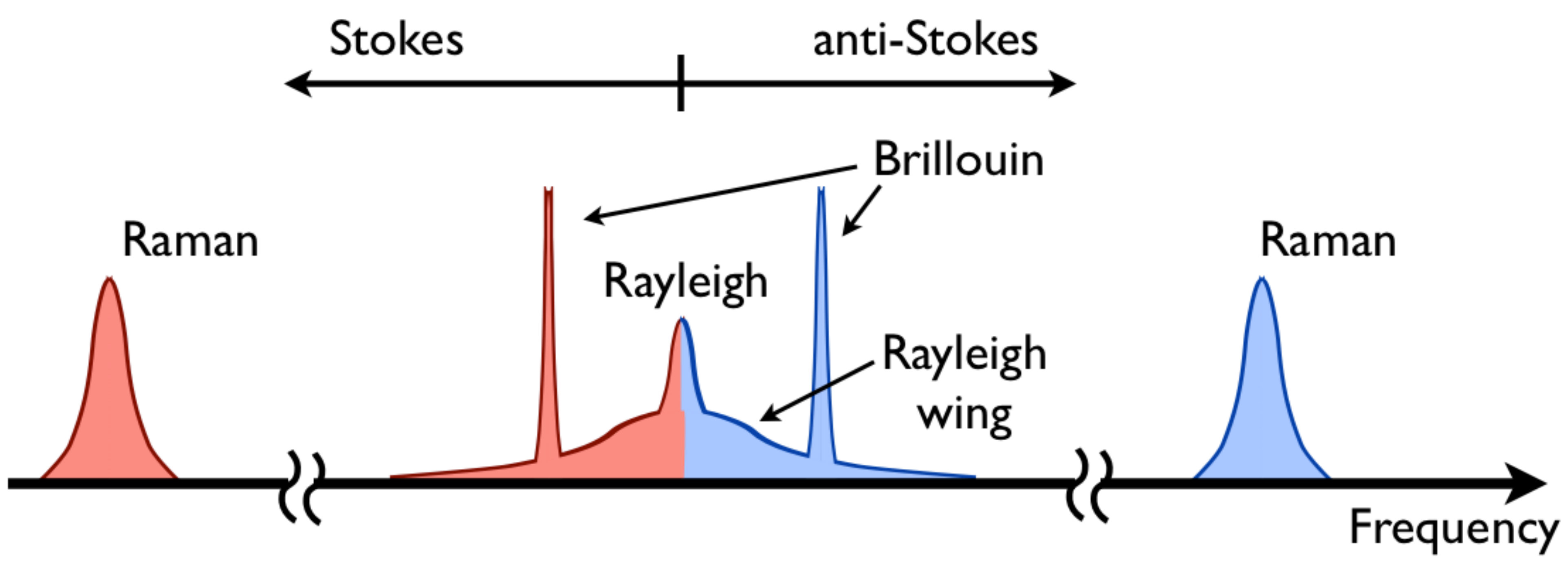}   %0.85   
		  }		
	\caption{ 
	\textbf{Concept:}
	 \textbf{(a)} Cooling of an acoustic wave through the interplay between anti-Stokes photoelastic scattering and electrostrictive pressure.
	 \textbf{(b) Selectively enabling Brillouin anti-Stokes cooling} (Fig. a), while rejecting Stokes heating is achieved by relying on a resonator with asymmetric optical resonance structure (right) that matches the energy and momentum of the acoustical mode being cooled (left).
	\textbf{(c)} 
	Pump photons photoelastically scatter from the Brownian acoustical mode, and thermal energy is removed since only the anti-Stokes process is resonantly enhanced. Inset: FEM calculation of an acoustical mode with 12 wavelengths around the circumference, corresponds to the frequency $\Omega_a = 95$ MHz on a 77.5 {\textmu}m radius sphere. The amplitude and index perturbation are calculated for 300 K.
	\textbf{(d)} The family of material-level scattering processes (after \cite{Boyd}).
	\label{fig:Theory}}
	\end{adjustwidth}
\end{figure}

Spontaneous Raman- and Brillouin-scattering are common to almost any media.
Incident photons are annihilated in these processes, giving rise to scattered photons at redder Stokes or bluer anti-Stokes frequencies (see Figure~\ref{fig:Theory}d). These scattering events lead to the creation or annihilation of phonons respectively. 
While prior research in optomechanics has used forces including centrifugal radiation pressure \cite{Carmon2005,Arcizet:2006p1092, Gigan:2006p1091, Kleckner:2006p1082, Chan_GroundState_2011, Riviere:2011cj} and optical gradient force \cite{Povinelli:2005tv,Li:2008jb,Eichenfield:2009gj} to excite mechanical motion, it is only recently that Brillouin scattering and electrostriction \cite{GrudininCaF2lasing,Tomes2009,Bahl:2011cf} were demonstrated in microcavities.
Further, extensive work in radiation pressure cooling \cite{Chan_GroundState_2011, Riviere:2011cj, Arcizet:2006p1092, Gigan:2006p1091, Kleckner:2006p1082} raises the question of whether it is possible to 
cool phonon modes via a Brillouin process, similar to the scattering process available in bulk media \cite{Boyd}.
For achieving Brillouin anti-Stokes cooling the heating Stokes line needs to be eliminated.
However, since low acoustic frequencies separate the pump line from the Stokes and anti-Stokes lines, filtering out the Stokes line against the anti-Stokes line requires a rapid transmission change over an extremely small frequency difference, which is not easily available in bulk materials. Here we transform the energy flow direction in spontaneous Brillouin scattering to enable annihilation of phonons and reduction of the effective acoustic mode temperature. 
Our technique for breaking the material heating-cooling symmetry relies on an ultra-high $Q$ microresonator with an asymmetric spectrum of resonances. 
Cooling is mediated by a pump optical wave, and a scattered optical wave, that beat together and electrostrictively attenuate the Brownian motion of the mechanical mode.
\\

\begin{figure}[h]
	\centering
		\includegraphics[width=\hsize]{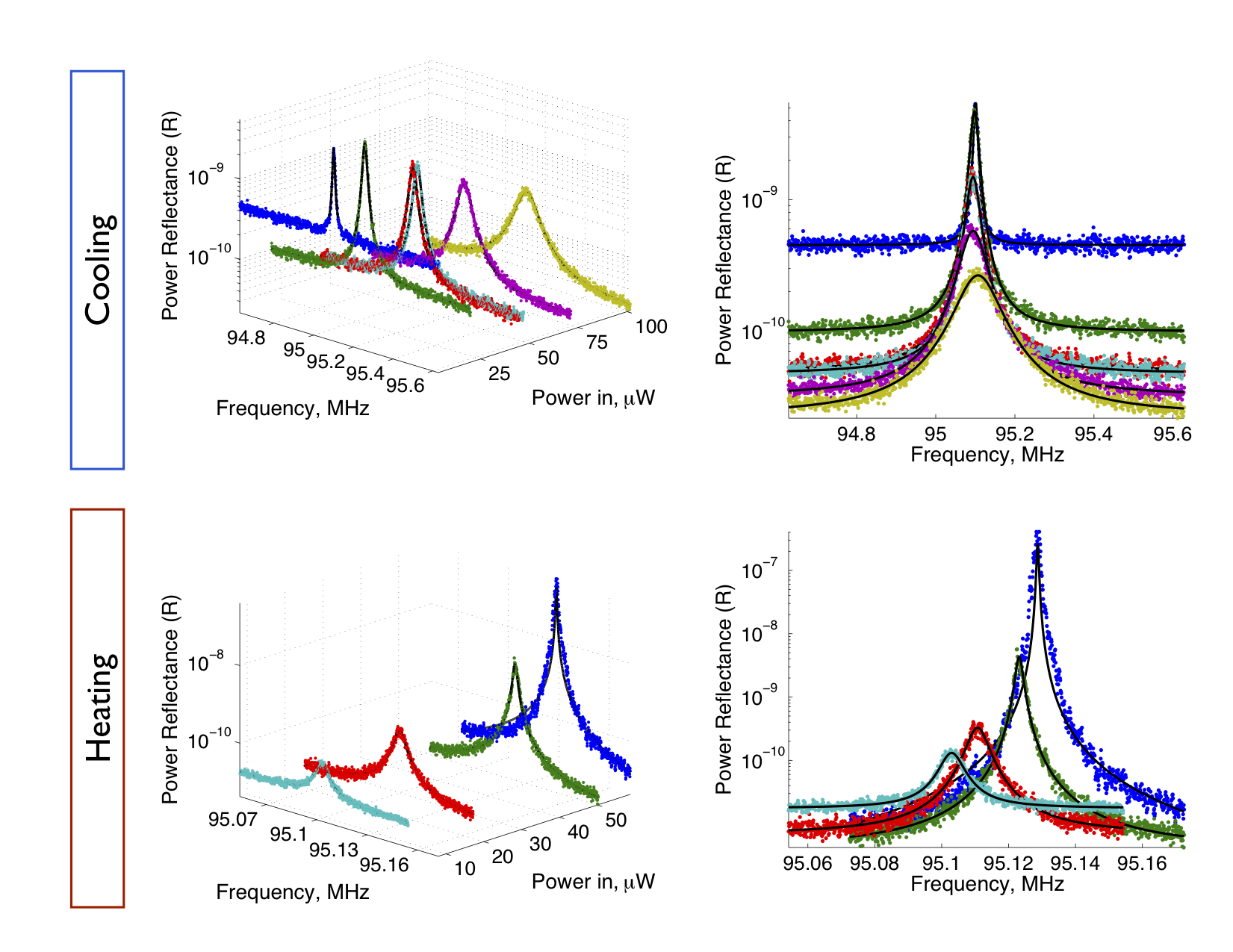}
	\caption{\textbf{Experimental observation of cooling} and heating of a 95 MHz acoustical mode in the microsphere system. The beating spectra between the pump and scattered optical signals are presented here, which are a measurement of the acoustical mode. 
	\textbf{Cooling is observed} in the broadening of this spectrum as a function of increasing pump power (top).
	To cool the acoustical mode, we pump the low frequency optical resonance ($O_P$) in order to only resonantly enhance the anti-Stokes process while suppressing the Stokes process. For the heating experiment (bottom), we pump the higher frequency optical resonance ($O_{aS}$) instead, to resonantly enhance the Stokes process.
	\label{fig:Experiment}}
\end{figure}

\begin{figure}[p]  %[h!]
	\centering
		\subfigure[] { 
			\includegraphics[width=0.55\hsize, clip=true, trim=0.5in 0.6in 5.2in 0.4in]{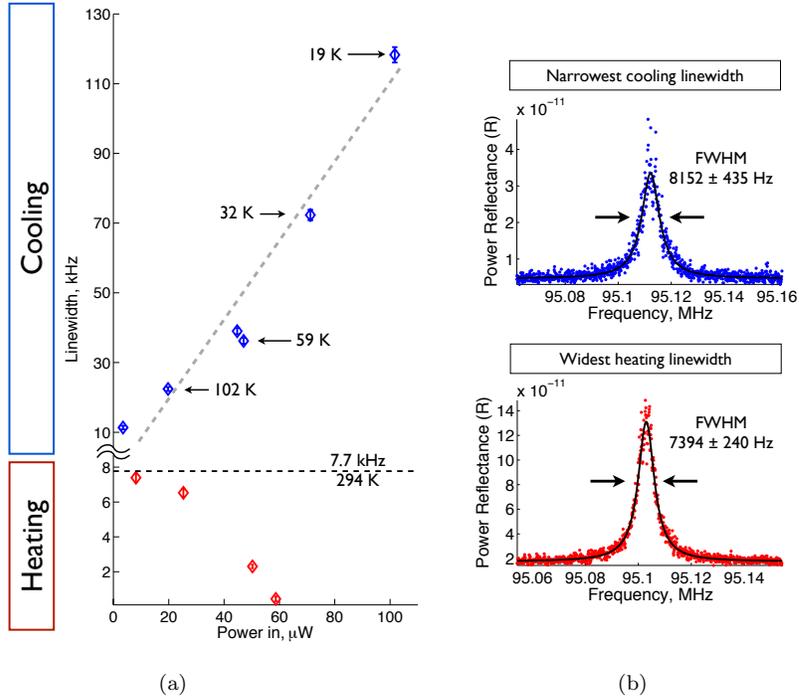}  %3.8in bottom
		}
		\subfigure[] { 
			\includegraphics[width=0.40\hsize, clip=true, trim=6in 0.6in 1in 0.4in]{Combined_HeatCool_LW_5Lor.pdf}  %3.8in bottom
		}	
		\caption{\textbf{Experimental result:} 
	\textbf{(a)} \textbf{We capture linewidth data} from a cooling and heating experiment, and calculate the effective acoustical mode temperatures. As expected, both cooling and heating linewidths converge towards the Brownian room temperature linewidth at low optical pump power. Dashed line is a guide for the eye.
	\textbf{(b)} We calibrate the acoustical linewidth as a function of temperature by finding the point of convergence of the cooling and heating data. The narrowest observed cooling linewidth (top) and the widest observed heating linewidth (bottom) are averaged to estimate the 7.7~kHz linewidth for the acoustical mode at ambient temperature of 294~K.
	\label{fig:Linewidth}}
\end{figure}

\begin{figure}[p]  %[h!]
	\centering
		\subfigure[] { 
			\includegraphics[width=0.75\hsize, clip=true, trim=1.6in 2.9in 2.2in 0.5in]{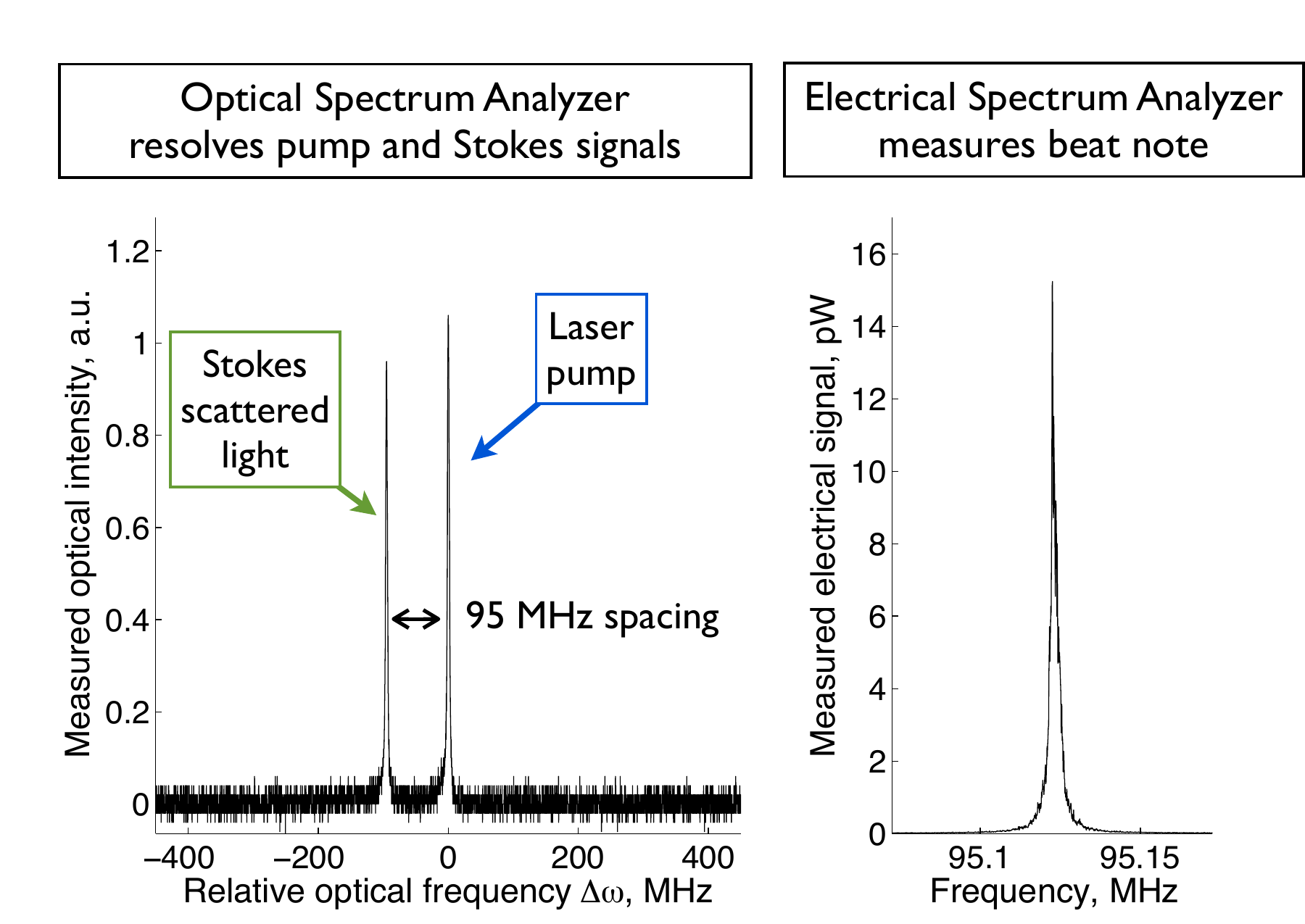}  %3.8in bottom
			%1.6in 2.9in 2.2in 0in
		}
		\subfigure[] {
			\includegraphics[width=0.75\hsize, clip=true, trim=1.6in 0.4in 2.2in 5.2in]{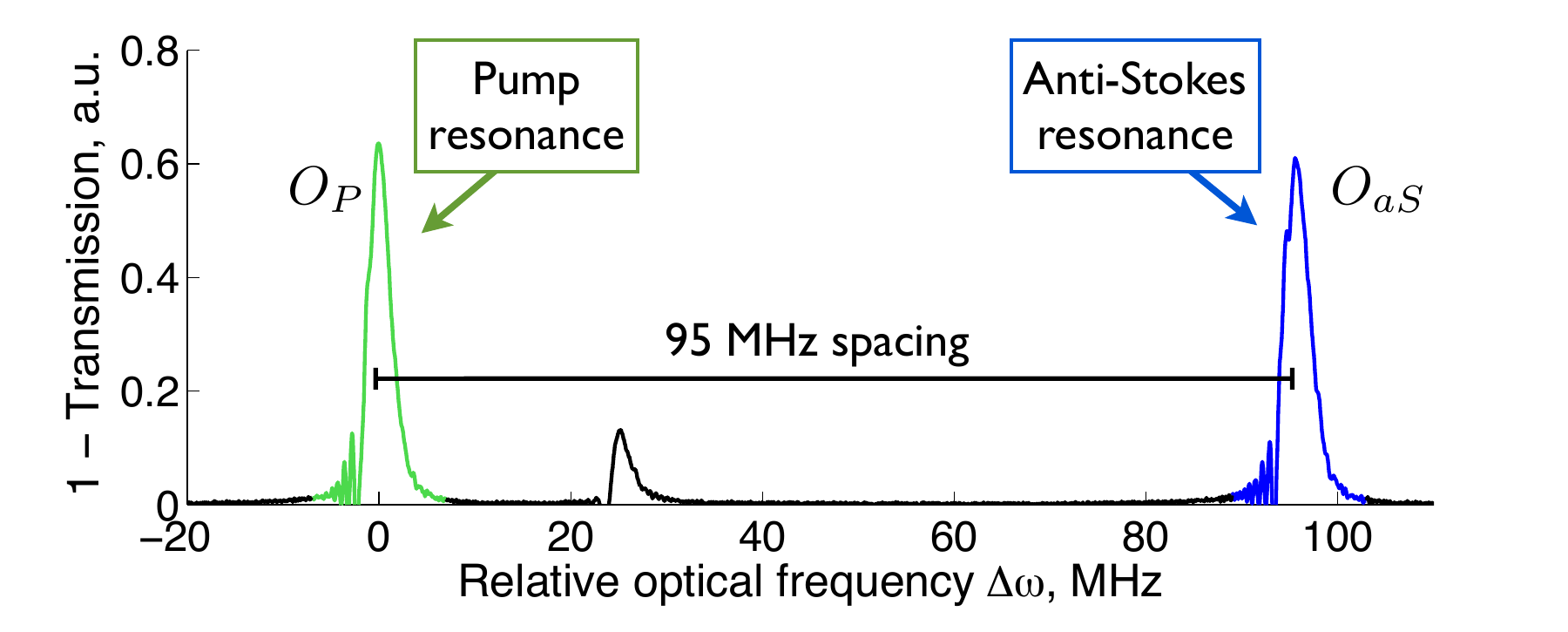}
		}
	\caption{\textbf{The participating resonances are analyzed via three independent experimental measurements}, confirming that 
	\textbf{(a)} the optical signals beating on the photodetector (right) indeed originate from one pump and one Stokes optical line as resolved with an optical spectrum analyzer (left). 
	\textbf{(b)} The separation between these optical lines corresponds with the transmission spectrum of our device, measured by sweeping the pump wavelength. Absorption peaks that overlap the lines resolved in (a, left) are seen. Resonances are named according to their role in the cooling experiments.
	\label{fig:ExptPhaseMatch}}
\end{figure}

Our experiment (Figure~\ref{fig:Theory}c) is based on a silica microsphere resonator \cite{GorodetskyMicrospheres} of $Q_{\textrm{optical}} > 10^8$.
We evanescently couple light into the resonator via a tapered optical fiber \cite{Knight:1997p1319}.
%Cai:01, Spillane:2003p1318, Knight:1997p1319}. 
Light is coupled out from the other side of this taper, allowing interrogation of the acoustical mode via the light that is scattered from it.
This spherical resonator supports the three actors that participate in the Brillouin cooling process -- two optical resonances (pump and anti-Stokes) and one acoustical resonance. The acoustical as well as the optical modes are of a whispering-gallery type and circulate in unison with considerable overlap.
These three modes are coupled via an interplay of photoelastic scattering and optical electrostriction (Figure \ref{fig:Theory}a), which in our case allows cooling of the acoustical mode. 
Experimental interrogation of the acoustical mode is made possible through observations of the pump light that it scatters (and Doppler shifts) into the anti-Stokes mode. The beat note between the pump light and the anti-Stokes light is measured on a photodetector at the output of the taper and is a measurement of the acoustical mode.
The mechanical deformation of the sphere is illustrated with 12 acoustical wavelengths along the circumference in Figure~\ref{fig:Theory}c. This illustration corresponds to an experimentally observed (see Figures \ref{fig:Experiment}--\ref{fig:ExptPhaseMatch}) acoustical mode of frequency $\Omega_a = 95$ MHz on a 77.5 {\textmu}m radius sphere, which we numerically calculate in the Supplement. In this work, we will experimentally cool this mechanical mode.
\\

While energy considerations give a general explanation for cooling, an explicit solution is needed. We therefore also analytically derive the cooling process while starting from the acoustical- and optical-wave equations in the Supplement. The source terms in these equations are taken to be electrostrictive pressure for the acoustical wave, and photoelastic index change for the optical wave. As expected from energy considerations, the solution shows that light applies compressive pressure on the region of the acoustical wave that expands, to take away energy from sound. 
We also confirm that repeating the same calculation for the Stokes process gives pressure on the acoustical regions while they shrink so that energy is imparted to the acoustical wave as expected. 
The classical analysis also shows that higher acoustical and optical quality factors, and lower acoustical frequencies, improve the cooling ratio (Eq. S.44).
A quantum analysis \cite{Tomes_coolingTheory} provides more details on the feasibility of this system towards ground-state cooling \cite{Chan_GroundState_2011,Riviere:2011cj,Teufel:2011jg}.
In what follows we will explain our technique that enables anti-Stokes cooling (Figure \ref{fig:Theory}a), without the degradation caused by the heating Stokes process. 
\\

The major enablers for Brillouin cooling are (i) the \textbf{selective resonant enhancement} of the anti-Stokes cooling process over the Stokes process, and (ii) \textbf{phase matching} for the two optical modes and the acoustic mode interacting through the Brillouin process.
Selective resonant enhancement of the anti-Stokes process (Figure \ref{fig:Theory}a) is made possible due to the non-periodic frequency separation \cite{PhysRevA.76.023816,PhysRevLett.100.103905} of the high order optical resonances of the sphere. 
Here we exploit this non-periodic resonance structure for solely selecting the anti-Stokes line.
The phase matching requirement implies that the anti-Stokes photon should carry out the energy ($\omega$) and momentum ($M$) of the pump photon and the thermal phonon. In a spherical resonator, the momentum parameter $M$ relates to azimuthal propagation $e^{i(M \, \phi \, - \, \omega \, t)}$ around the sphere equator. 
As expected, these conservation considerations are in agreement with our analytically derived synchronous solution for the coupled wave equations \cite{Boyd} that describe our system (See Supplement). 
Both phase matching and selective resonant enhancement are illustrated in the energy-momentum ($\omega$--$M$) diagram of Figure \ref{fig:Theory}b. Since the pump optical resonance ($\omega_P$,~$M_P$) and the anti-Stokes resonance ($\omega_{aS}$,~$M_{aS}$) are separated by the acoustical resonance parameters ($\Omega_a$,~$M_a$), as indicated by the solid triangles, both phase match and resonant enhancement are satisfied for anti-Stokes scattering.
On the other hand, the dashed triangle of Figure \ref{fig:Theory}b illustrates the off-resonantly eliminated Stokes process, where subtraction of the same acoustical parameters ($\Omega_a$,~$M_a$) from the pump resonance ($\omega_P$,~$M_P$) brings us to a region ($\omega_S$,~$M_S$) where no optical resonances exist.
\\

\textit{Experimental measurement of Brillouin cooling} is shown in Figure \ref{fig:Experiment}, as indicated by the broadening of the beating signal between pump and anti-Stokes lines, 
as a function of increasing pump power. This beat observed from the tapered fiber coupler serves as a measurement of the acoustical mode (See Supplement).
The cooling experiment is performed by positioning the pump laser at the lower frequency resonance ($\omega_P$ in Figure \ref{fig:Theory}b), and by observing the light scattered in the anti-Stokes direction into the $\omega_{aS}$ resonance.
As in \cite{Metzger:2004p1357,Arcizet:2006p1092} we prefer linewidth as a measurement of cooling \cite{Riviere:2011cj,Teufel:2011jg,Chan_GroundState_2011,Metzger:2004p1357,Arcizet:2006p1092,Naik:2006gf,Gigan:2006p1091,Kleckner:2006p1082,Kippenberg29082008} since the alternative of measuring the integrated power of the beat note might be affected by attenuation of the optical signals.
Since the power reflectance spectrum is proportional to the squared displacement spectrum of the acoustical mode (See Supplement), the acoustical linewidth has been extracted from this data and is presented in Figure~\ref{fig:Linewidth}a.
Here, the linewidth of the 95~MHz acoustical mode increases from 7.7~kHz to 118~kHz with increasing pump power, in an open laboratory environment at 294~K. 
We measure the effective mode temperature through its inverse proportionality to the linewidth of the acoustical mode (See Supplement).
The result indicates an achieved cooling ratio of 15, implying an effective mode temperature of 19~K at the maximum observed cooling point.  
The cavity operates in the thermally stable regime \cite{CarmonThermal}. The cooling factor we achieved was limited by the optical power available from the pump laser.
To complete the analysis, we now invert this process by moving the pump to the higher frequency resonance. The low frequency optical mode now functions as a Stokes resonance for the higher frequency pump. As expected, line narrowing is now observed as indicated in Figure \ref{fig:Linewidth}a, accompanied by growth of the Stokes signal (Figure \ref{fig:Experiment}, bottom).
The Stokes process linewidth and the anti-Stokes cooling linewidth converge at low input powers, indicating the mechanical quality factor of the mode. This point of convergence provides the 7.7 kHz acoustical linewidth calibration at 294~K in Figure \ref{fig:Linewidth}, so that mode temperature can be calculated during the cooling experiment.
\\

\textit{We experimentally confirm that two optical resonances exist} that are separated by the mechanical frequency, as required for phase matched cooling. 
We already know that the pump to anti-Stokes beat, measured by a photodetector, provides us with a frequency of 95 MHz for the mechanical mode (Figure~\ref{fig:ExptPhaseMatch}a, right). We then perform the following two independent measurements that obtain no information from this electrical measurement.
Measuring resonator transmission while scanning the pump frequency (Figure~\ref{fig:ExptPhaseMatch}b) indeed confirms the existence of two optical resonances spaced by 95~MHz to within measurement error. 
We also verify that the relevant resonance structure is indeed asymmetric relative to the pump by confirming that no resonance appears on the Stokes side of pump resonance $O_P$ at a spacing of 95~MHz.
The optical line frequency spacing is also measured by repeating the Stokes experiment \cite{Bahl:2011cf} but now with an optical spectrum analyzer at the output.
As expected, the spectrum analyzer resolves two optical lines that are separated by 95 MHz (Figure~\ref{fig:ExptPhaseMatch}a, left) in agreement with the measured transmission profile.
These three independent measurements -- beat spectrum, optical transmission, and the optical output spectrum -- confirm the existence of the resonant triplet that was used in the cooling process.
\\

Here we use forward Brillouin scattering \cite{Shelby:1985p1169, Bahl:2011cf} in a system involving two optical modes and one acoustical mode to demonstrate cooling.  In comparison, previous optical cooling studies have used fluorescence \cite{EpsteinBook}, radiation pressure forces \cite{Chan_GroundState_2011, Riviere:2011cj, Arcizet:2006p1092, Gigan:2006p1091, Kleckner:2006p1082}, and photothermal pressure \cite{Metzger:2004p1357}.
While the end result here combines a net annihilation of phonons \cite{EpsteinBook,Chan_GroundState_2011, Teufel:2011jg,Riviere:2011cj, Arcizet:2006p1092, Gigan:2006p1091, Kleckner:2006p1082,Metzger:2004p1357} in a multi-resonance system \cite{Grudinin:2010fe,GrudininCaF2lasing,Tomes2009,Bahl:2011cf}, no detuning of the pump with respect to the resonance is required for achieving cooling. Additionally, the fact that the mechanical mode is of a whispering-gallery type here, suggests low dissipation via the support \cite{Bahl:2011cf}.
While we selected the 95 MHz mode in this study because of the greater cooling ratio achievable for such relatively low frequencies (Eq. S.44), mechanical modes spanning from tens of MHz \cite{Bahl:2011cf} to tens of GHz \cite{GrudininCaF2lasing,Tomes2009} can be accessed with this method.
In a broad context, Brillouin scattering belongs to a family of scattering processes that includes Raman- and Rayleigh-scattering (Figure~\ref{fig:Theory}d). Reversing the energy flow direction here, in respect with Brillouin systems such as \cite{Shelby:1985p1169, Bahl:2011cf, Tomes2009, GrudininCaF2lasing}, raises the question of whether similar inversion is possible with Raman scattering \cite{Kang:2009p1366} non-optomechanical cooling of solids \cite{EpsteinBook,VermeulenRaman,Jalali:2007p1364} in spite of the higher frequencies involved.

\section*{Acknowledgements}

This work was supported by the Defense Advanced Research Projects Agency (DARPA) Optical Radiation Cooling and Heating in Integrated Devices (ORCHID) program through a grant from the Air Force Office of Scientific Research (AFOSR). M.T. is supported by a National Science Foundation fellowship. F.M. acknowledges the Emmy-Noether program.

%%%%%%%%%%%%%%%%%%%%%%%%%%%%%%%%%
%\bibliographystyle{ieeetr}
\bibliographystyle{unsrt}
\bibliography{BrillCool}
%\input{Cooling14_ArXiv_Combined.bbl}
%%%%%%%%%%%%%%%%%%%%%%%%%%%%%%%%%

\newpage

\clearpage

\title{Supplementary information\\Brillouin Cooling}
\date{}
\maketitle

\section{Numerical calculation of acoustical mode and photoelastic grating}
\label{sec:Simulation}

\begin{figure}[b!]
	\centering
		\includegraphics[width=0.9\hsize]{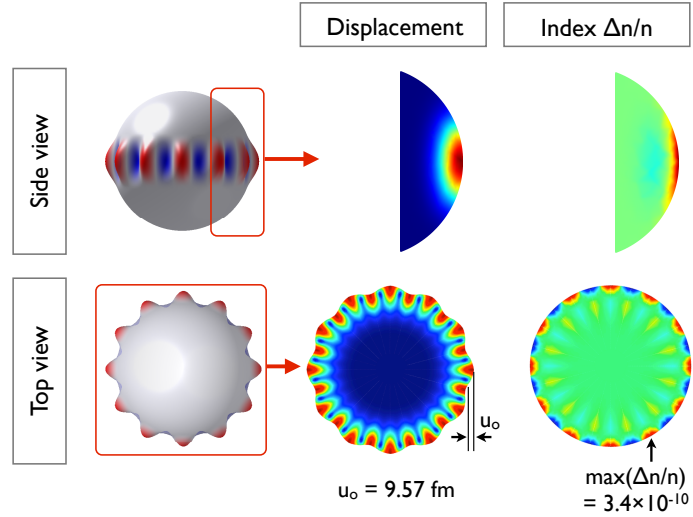}
	\caption{\textbf{Finite-element calculation:} The $M_a=12$, $\Omega_a = 95$ MHz mechanical mode is numerically calculated. The mechanical amplitude is calculated to be 9.57 fm as a result of Brownian phonon occupation of the mode at room temeprature. The photoelastic grating ($\Delta n/n$) is then numerically calculated using the effect of strain.
	\label{fig:Mode}}
\end{figure}

A question arises on the photoelastic mechanism that turns strain and deformation, associated with the acoustical mode, into a traveling index perturbation that scatters and Doppler shifts light. Two effects must be considered in this calculation -- (i) the strain-induced change of refractive index \citeS{Ilchenko:1998p1359}, and (ii) the shifting of the silica-air boundary as a function of the deformation \citeS{PhysRevE.65.066611}. 

Finite-element simulations (Figure~\ref{fig:Mode}) are used to numerically calculate a surface mode at the experimentally observed acoustical frequency of 95 MHz.
We then equate the numerically calculated strain energy of the acoustical mode to $k_B \, T$, to determine the mode amplitude at room temperature. We have determined that our $M_a=12$, $\Omega_a = 95$ MHz acoustical mode has a Brownian amplitude $u_o = 9.57~\textrm{fm}$ at room temperature (see section \ref{sec:BrownianAmplitude}). 

In the second stage of this calculation, the strain and deformation outputs of the numerical solver are used to calculate the index change using \citeS{Ilchenko:1998p1359}, \citeS{PhysRevE.65.066611}. Accordingly, the refractive-index perturbation of the Brownian photoelastic grating is estimated to have a maximum amplitude of $\Delta n/n = 3.4 \times 10^{-10}$.

\section{Analytical derivation of anti-Stokes Brillouin cooling}

\subsection{Problem setup}

Here we will use coupled-mode theory \citeS{HausCoupledTheory} to couple two optical resonances and one mechanical resonance using the optical wave equation and the acoustical wave equation. Our analysis assumes two optical modes with similar polarization, which allows for considering only the scalar value of the field.

\begin{figure}[tb]
	\centering
		\includegraphics[width=0.7\hsize, clip=true, trim=1in 3in 1.1in 2in]{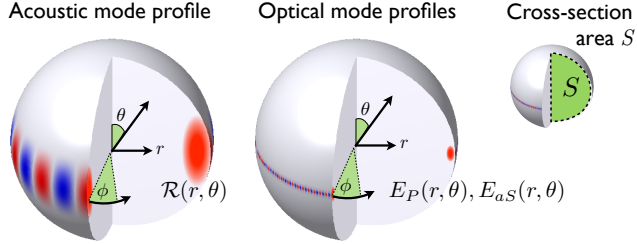}
	\caption{\textbf{Illustration of spatial mode profiles:} The cross-sectional acoustic mode profile is represented as $\mathcal{R}(r,\theta)$, while the profiles for the pump and anti-Stokes optical modes can be represented as $E_P(r,\theta)$ and $E_{aS}(r,\theta)$. We define the cross-sectional area transverse to propagation as $S$.
	\label{fig:ModeProfiles}}
\end{figure}

Consider two propagating optical fields $\tilde{E_P}$ and $\tilde{E_{aS}}$ within the spherical resonator representing `Pump' and `anti-Stokes' modes respectively. The fields are defined in a spherical coordinate system ($r, \theta, \phi$) as
\begin{align}
	\tilde{E_P} & = A_P(t) ~ E_P(r,\theta) ~ \cos(M_P ~ \phi - \omega_P ~ t)  \label{eqn:TEP} \\
	\tilde{E_{aS}} & = A_{aS}(t) ~E_{aS}(r,\theta) ~\cos(M_{aS} ~ \phi - \omega_{aS} ~ t + \psi)   \label{eqn:TEaS}
\end{align}
where $A_P(t)$, $A_{aS}(t)$ are the real and positive amplitudes of the waves, and $E_P(r,\theta)$, $E_{aS}(r,\theta)$ are the field distributions in the plane normal to propagation (see Figure~\ref{fig:ModeProfiles}). The fields propagate in the azimuthal direction ($\phi$) with integer propagation constants of $M_P$ and $M_{aS}$, and optical frequencies $\omega_P$ and $\omega_{aS}$. The anti-Stokes field is assigned a relative phase shift $\psi$.
Later on, this phase $\psi$ will be explicitly calculated in order to satisfy a synchronous solution of the coupled wave equations. This phase will turn out to mathematically express the major deliverable in this paper as it dictates the energy flow direction from sound to light (and vice versa if $\omega_{aS}$ is replaced with Stokes frequency).
The sinusoidal representations are chosen instead of phasors for mathematical convenience. 

The acoustic density wave $\tilde\rho$ is also defined without the limitation of generality as
\begin{align}
	\tilde{\rho} & = \rho_o + \Delta\tilde\rho  \notag \\
		& = \rho_o +  \rho(t)~ \mathcal{R}(r,\theta)~ \cos( M_a ~ \phi - \Omega_a ~ t)
	\label{eqn:AcousticWave}
\end{align}
where $\rho_o$ is the unperturbed density of the material, $\rho(t)$ is the amplitude of the wave, $\mathcal{R}(r,\theta)$ spatial density distribution, $M_a$ is the acoustic propagation constant, and $\Omega_a$ is the acoustic frequency.

It is convenient to choose normalized transverse profiles for the three fields, such that
\begin{align}
	& \iint_S ~ \mathcal{R}(r,\theta)^2 ~ dS = 1 \label{eqn:NormR} \\
	& \iint_S ~ E_P(r,\theta)^2 ~ dS = 1  \label{eqn:NormP} \\
	& \iint_S ~ E_{aS}(r,\theta)^2 ~ dS = 1  \label{eqn:NormAS}
\end{align}
where $S$ is the cross-sectional area shown cut away in Figure \ref{fig:ModeProfiles}.

\subsection{Optical wave equation with a photoelastic scattering source}
\label{sec:Optical}

Consider the electromagnetic wave equation
\begin{align}
	\nabla^2  \tilde{E} & = \mu_o \frac{\partial^2}{\partial t^2} \tilde{D}  \notag \\
		& = \mu_o \frac{\partial^2}{\partial t^2} \left( \epsilon_o \epsilon_r \tilde{E} \right)  \notag \\
		& \approx \mu_o \epsilon_o \frac{\partial^2}{\partial t^2} \left((n_o^2 + 2 n_o \Delta\tilde{n})  \tilde{E}  \right)
\end{align}
where we have discarded the higher order terms in $\epsilon_r = (n_o + \Delta\tilde{n})^2$. Here, $n_o$ is the refractive index of the material, and $\Delta\tilde{n} \sim \cos( M_a ~ \phi - \Omega_a ~ t)$ is the periodic modulation photoelastically applied to the index through the acoustic density wave. 
The terms can then be rearranged to obtain
\begin{align}
	\nabla^2  \tilde{E} - \mu_o \epsilon_o n_o^2 \frac{\partial^2}{\partial t^2} \tilde{E} = \mu_o \epsilon_o \frac{\partial^2}{\partial t^2} \left( 2 n_o \Delta\tilde{n} ~ \tilde{E}  \right)
\end{align}
In our case, $\tilde{E} = \tilde{E_P} + \tilde{E_{aS}}$ is the total electric field. The right hand side of this equation is a source term that can be rewritten in the form of photoelastic scattering of the optical fields $\tilde{E}$ from the density wave $\Delta\tilde\rho$ as follows.
\begin{align}
	\label{eqn:OpticalWave}
	\nabla^2  \tilde{E} - \frac{n_o^2}{c^2} \cdot \frac{\partial^2 {\tilde{E}}}{\partial t^2} = \frac{\mu_o \epsilon_o \gamma_e}{\rho_o} \cdot \frac{\partial^2}{\partial t^2}  \left( \Delta\tilde\rho ~ \tilde{E} \right)
\end{align}
where we have used the relations 
\begin{align}
	& \Delta\tilde\epsilon_r = 2 n_o ~ \Delta\tilde{n} = \left( \frac{\partial \epsilon}{\partial \rho} \right) \Delta\tilde\rho   \label{eqn:relate_index_density} \\
	& \gamma_e = \rho_o ~ (\partial \epsilon / \partial \rho) ~.
\end{align}
The electrostrictive constant $\gamma_e$ is a material property governed by the material density $\rho_o$ and the change of permittivity with change in density (See \citeS{Boyd}, chapter 9). 
\\

The optical wave equation \ref{eqn:OpticalWave} can be approached via coupled-mode theory \citeS{HausCoupledTheory} to determine a synchronous solution for the anti-Stokes wave $\tilde{E_{aS}}$. For this, we only consider the synchronous terms at frequency $\omega_{aS}$. Equation \ref{eqn:OpticalWave} is then reduced to
\begin{align}
	\label{eqn:OpticalWave2}
	\nabla^2  \tilde{E_{aS}} - \frac{n_o^2}{c^2} \cdot \frac{\partial^2 {\tilde{E_{aS}}}}{\partial t^2} = \frac{\mu_o \epsilon_o \gamma_e}{\rho_o} \cdot \frac{\partial^2}{\partial t^2}  \left( \Delta\tilde\rho ~ \tilde{E_P} \right) ~.
\end{align}
Equation \ref{eqn:OpticalWave2} shows that the coexistence of the Brownian density fluctuation \ref{eqn:AcousticWave} and the pump optical mode \ref{eqn:TEP} constitute an optical source term that shines light into the optical anti-Stokes mode \ref{eqn:TEaS}. This mathematical result describes, as expected, an experimental reality where the Brownian density fluctuation scatters the pump mode into the anti-Stokes mode while properly Doppler shifting its color.
\\

The L.H.S. of \ref{eqn:OpticalWave2} can then be simplified to
\begin{align}
	\nabla^2  \tilde{E_{aS}} ~ - ~ \frac{n_o^2}{c^2} & \cdot \frac{\partial^2 {\tilde{E_{aS}}}}{\partial t^2} \notag \\
		 = ~ & A_{aS}(t)  \left( \nabla^2 -\frac{n_o^2}{c^2} \right)  E_{aS}(r,\theta) \cos(M_{aS} ~ \phi - \omega_{aS} ~ t + \psi) \notag \\
	& -\frac{n_o^2}{c^2} ~ 2 ~ \omega_{aS} A'_{aS}(t) E_{aS}(r,\theta) \sin(M_{aS} ~ \phi - \omega_{aS} ~ t + \psi)
	\label{eqn:OpticalLHS1}
\end{align}
where we make the slowly-varying amplitude approximation $A''_{aS}(t) \approx 0$. 
The first term on the R.H.S. of equation \ref{eqn:OpticalLHS1} equals zero as it is the unperturbed solution for the anti-Stokes mode.

We further make the simplification that regions of non-zero optical intensity are concentrated at a radial location $\mathfrak{R}_{opt}$ that is almost equal to the sphere radius (see Figure \ref{fig:ModeProfiles}). 
This simplification is acceptable since the transverse profile of the optical mode is of the order of 1 {\textmu}m \citeS{Oxborrow} while the sphere radius is of the order of 100 {\textmu}m.
Equation \ref{eqn:OpticalLHS1} then becomes
\begin{align}
	\nabla^2  \tilde{E_{aS}} - \frac{n_o^2}{c^2} \cdot \frac{\partial^2 {\tilde{E_{aS}}}}{\partial t^2}  \approx - \frac{2 M_{aS}^2}{\mathfrak{R}_{opt}^2 \omega_{aS}}  A'_{aS}(t) E_{aS}(r,\theta) \sin(M_{aS} ~ \phi - \omega_{aS} ~ t + \psi)
	\label{eqn:LHSOptical}
\end{align}
where the relation $c/n = \mathfrak{R}_{opt} ~ \omega_{aS} / M_{aS}$ has been used.
\\
\\

Considering the R.H.S. of \ref{eqn:OpticalWave2}, the product term of $\Delta\tilde\rho$ and $\tilde{E_P}$ provides the simplification
\begin{align}
	 \Delta\tilde\rho ~ \tilde{E_P} 
	 	& = \rho(t) \mathcal{R}(r,\theta)A_P(t) E_P(r,\theta) 
	 		\cdot \cos(M_a \, \phi - \Omega_a \, t)  \cdot  \cos(M_P \, \phi - \omega_P \, t)  \notag \\
		& = \rho(t) \mathcal{R}(r,\theta)A_P(t) E_P(r,\theta) 
	 		 \cdot \frac{1}{2} \cos((M_P + M_a) \phi - (\omega_P + \Omega_a) t)
\end{align}
where in the last step only the term providing for anti-Stokes scattering is preserved.
In order to satisfy a synchronous solution to \ref{eqn:OpticalWave2}, the following two relations appear
\begin{align}
	& \omega_{aS} = \omega_P + \Omega_a \label{eqn:ConservationW}\\
	& M_{aS} = M_P + M_a ~ .  \label{eqn:ConservationM}
\end{align}
This result is expected from energy and momentum conservation considerations. Equations \ref{eqn:ConservationW} and \ref{eqn:ConservationM} describe a situation where the created anti-Stokes photon carries the momentum and energy of the annihilated thermal phonon and pump photon. 
This result is applied to the R.H.S. of \ref{eqn:OpticalWave2} to give
\begin{align}
	& \frac{\mu_o \epsilon_o \gamma_e}{\rho_o} \cdot %\left. 
	\frac{\partial^2}{\partial t^2}  \left( \Delta\tilde\rho ~ \tilde{E_P} \right) \notag \\
	& = \frac{\mu_o \epsilon_o \gamma_e}{\rho_o}  \rho(t) \mathcal{R}(r,\theta)A_P(t) E_P(r,\theta) \cdot \frac{1}{2} \frac{\partial^2}{\partial t^2} \cos((M_P + M_a) \phi - (\omega_P + \Omega_a) t) \notag \\
	& = - \frac{\mu_o \epsilon_o \gamma_e}{2 \rho_o}  \rho(t) \mathcal{R}(r,\theta)A_P(t) E_P(r,\theta) ~ \omega_{aS}^2 ~ \cos(M_{aS} ~ \phi - \omega_{aS} ~ t) ~.
	\label{eqn:RHSOptical}
\end{align}
where the slowly-varying amplitude approximation has been applied.
\\

Combining the L.H.S. and R.H.S. of \ref{eqn:OpticalWave2} (from \ref{eqn:LHSOptical} and \ref{eqn:RHSOptical}), multiplying both sides by $E_{aS}(r,\theta)$, and integrating over the area $S$, we obtain
\begin{align}
	 - & \frac{2 M_{aS}^2}{\mathfrak{R}_{opt}^2 \omega_{aS}}  A'_{aS}(t) \sin(M_{aS} ~ \phi - \omega_{aS} ~ t + \psi) ~ \iint_S E_{aS}(r,\theta)^2 ~ dS \notag \\
	& = - \frac{\mu_o \epsilon_o \gamma_e}{2 \rho_o}  \rho(t) A_P(t) ~ \omega_{aS}^2 \cos(M_{aS} ~ \phi - \omega_{aS} ~ t) ~ \iint_S \mathcal{R}(r,\theta) E_P(r,\theta) E_{aS}(r,\theta) ~ dS
\end{align}
The normalization of equation \ref{eqn:NormAS} is then applied, and we reach the result
\begin{align}
	A'_{aS}(t) \, \sin(& M_{aS} ~ \phi - \omega_{aS} ~ t + \psi) \notag \\
		& =  C_1 ~ \rho(t) \, \cos(M_{aS} ~ \phi - \omega_{aS} ~ t) ~ \iint_S \mathcal{R}(r,\theta) E_P(r,\theta) E_{aS}(r,\theta) ~ dS
	\label{eqn:Aas_prime}
\end{align}
where $C_1$ is the positive pre-factor
\begin{align}
	C_1 = \frac{\mu_o \epsilon_o \gamma_e ~ \omega_{aS}^3 ~ \mathfrak{R}_{opt}^2 } { 4 ~ \rho_o ~ M_{aS}^2 } \cdot A_P(t) ~.
\end{align}
Without loss of generality, we assume that the modal overlap integral
\begin{align}
	\mathcal{I}_{\textrm{overlap}} = \iint_S   \mathcal{R}(r,\theta) E_P(r,\theta) E_{aS}(r,\theta) ~ dS
\end{align}
is a positive number. A synchronous solution for equation \ref{eqn:Aas_prime} (and the optical wave equation \ref{eqn:OpticalWave2}) is then obtained for the phase angle $\psi = \pi/2$. The simplified form of \ref{eqn:Aas_prime} is then
\begin{align}
	A'_{aS}(t) = C_1 ~ \rho(t) ~ \mathcal{I}_{\textrm{overlap}}
	\label{eqn:Aas_prime_simple}
\end{align}
\\

\subsection{Acoustic wave equation with electrostrictive pressure sources}
\label{sec:Acoustic}

The acoustic wave equation can be written as
\begin{align}
	\frac{\partial^2 {\tilde\rho}}{\partial t^2} -  \left( \frac{r \Omega_a}{M_a}\right) ^2 \nabla^2 \tilde\rho = \nabla^2 \tilde p_{st}
	\label{eqn:AcousticWaveEqn}
\end{align}
where the term within parentheses is the 
%radius $r$-dependent 
velocity of sound, and 
\begin{align}
	\tilde p_{st} = - \frac{1}{2} \epsilon_o \gamma_e | \tilde{E_P} + \tilde{E_{aS}} |^2
\end{align}
is the electrostrictive pressure generated by the total electric field (See \citeS{Boyd}, chapter 9). 
Similar to the treatment for the optical wave equation, the L.H.S. of the wave equation \ref{eqn:AcousticWaveEqn} can be simplified to
\begin{align}
	\label{eqn:AcousticLHS}
	\frac{\partial^2 {\tilde\rho}}{\partial t^2} -  \left( \frac{r \Omega_a}{M_a}\right) ^2 \nabla^2 \tilde\rho ~ \approx 2 ~ \Omega_a ~ \rho'(t) ~ \mathcal{R}(r,\theta) ~ \sin(M_a ~ \phi - \Omega_a ~ t)
\end{align}
where some terms have been removed through the use of the slowly-varying amplitude approximation,
as well as knowledge of the unperturbed acoustic wave equation.
\\

Considering the R.H.S. of the acoustic wave equation \ref{eqn:AcousticWaveEqn} we have
\begin{align}
	& | \tilde E_P  + \tilde E_{aS} |^2 \notag \\
		 & = |  A_P(t) E_P(r,\theta) \cos(M_P ~ \phi - \omega_P ~ t)  +  A_{aS}(t) E_{aS}(r,\theta) \cos(M_{aS} ~ \phi - \omega_{aS} ~ t + \psi)   |^2
\end{align}
which is simplified to
\begin{align}
	 | \tilde E_P  + \tilde E_{aS} |^2 
		\approx A_P(t) A_{aS}(t) & E_P(r,\theta) E_{aS}(r,\theta) \notag \\
		& \Bigl( \cos((M_{aS}-M_P) ~ \phi - (\omega_{aS}-\omega_P) t + \psi)   \notag \\
		& + \cos((M_{aS}+M_P) ~ \phi - (\omega_{aS}+\omega_P) t + \psi)  \Bigr)
		\label{eqn:Striction1}
\end{align}
where the high optical frequency terms at $2 \, \omega_P$ and $2 \, \omega_{aS}$, as well as the dc terms, have been discarded. Again, we employ the conservation relations \ref{eqn:ConservationW} and \ref{eqn:ConservationM} and preserve the synchronous term from \ref{eqn:Striction1} to match equation \ref{eqn:AcousticLHS}.
\begin{align}
	 \label{eqn:AcousticRHS}
	& \left. \tilde p_{st} \right|_{\textrm{at } \Omega_a} = \left. - \frac{1}{2} \epsilon_o \gamma_e  | \tilde E_P + \tilde E_{aS} |^2 \right|_{\textrm{at } \Omega_a} \\
		 & = - \frac{1}{2} \epsilon_o \gamma_e A_P(t) A_{aS}(t) E_P(r,\theta) E_{aS}(r,\theta) \cos((M_{aS}-M_P) ~ \phi - (\omega_{aS}-\omega_P) t + \psi) \notag \\
		 & = - \frac{1}{2} \epsilon_o \gamma_e A_P(t) A_{aS}(t) E_P(r,\theta) E_{aS}(r,\theta) \cos(M_a ~ \phi -\Omega_a ~ t + \psi) \notag
\end{align}
\\

The acoustical-source term (R.H.S. of \ref{eqn:AcousticWaveEqn}) is then evaluated as the Laplacian of the electrostrictive pressure using the spherical coordinate system.
\begin{align}
	\nabla^2 \tilde p_{st} =  \left( \frac{1}{r^2} \partial_r (r^2 \, \partial_r) + \frac{1}{r^2 \, \sin{\theta}} \partial_\theta (\sin{\theta} \, \partial_\theta) + \frac{1}{r^2 \, \sin^2{\theta}} \partial^2_\phi \right) \tilde p_{st}
	\label{eqn:SphLaplacian}
\end{align}
Although the common terminology for this term is the `source', it will be shown here to actually attenuate the Brownian density fluctuation. The R.H.S. of \ref{eqn:SphLaplacian} yields several terms including spatial derivatives of the electric field distribution. 

We now reassemble the two sides of the acoustic wave equation (\ref{eqn:AcousticWaveEqn}) from \ref{eqn:AcousticLHS} and \ref{eqn:SphLaplacian}, multiply both sides by $\mathcal{R}(r,\theta)$, and integrate over the area $S$ described in Figure \ref{fig:ModeProfiles}. Using the normalization for $\mathcal{R}(r,\theta)$ shown in \ref{eqn:NormR}, this gives us the equation
\begin{align}
	\label{eqn:rhodot1}
	2 \, \Omega_a \, \rho'(t) \sin(M_a ~ \phi - \Omega_a ~ t) ~ = ~ \iint_S \left( \nabla^2 \tilde p_{st} \right)  \, \mathcal{R}(r,\theta) ~ dS  ~.
\end{align}
Assuming that the optical modes are confined to the equatorial region as shown in Figure \ref{fig:ModeProfiles}, 
we have numerically determined that the last term of the Laplacian in \ref{eqn:SphLaplacian}, i.e.\ $(1 / r^2 \, \sin^2{\theta}) \partial^2_\phi$, is dominant in the solution of the R.H.S. integral in \ref{eqn:rhodot1}. 
We therefore obtain
\begin{align}
	\label{eqn:rhodot}
	& 2 \, \Omega_a \, \rho'(t) \sin(M_a ~ \phi - \Omega_a ~ t) \notag \\
	 	& \approx \frac{\epsilon_o \gamma_e A_P(t) A_{aS}(t) M_a^2}{2}  \cdot \cos(M_a ~ \phi -\Omega ~ t + \psi)  \cdot  \iint_S  \frac{E_P(r,\theta) E_{aS}(r,\theta)  \mathcal{R}(r,\theta)}{r^2 \, \sin^2 \theta}~ dS
\end{align}
Next, we apply the approximations $r \approx \mathfrak{R}_{opt}$ and $\theta \approx \pi/2$ for the denominator of the integral, since the integral only obtains non-zero values where the optical modes are present. 
We therefore have
\begin{align}
	\label{eqn:rhodot4}
	\rho'(t) 
	& \approx \frac{C_2 ~ A_{aS}(t)}{\mathfrak{R}_{opt}^2}  \cdot \frac{ \cos(M_a ~ \phi -\Omega ~ t + \psi)}{\sin(M_a ~ \phi - \Omega_a ~ t) }  \cdot \iint_S  E_P(r,\theta) E_{aS}(r,\theta)  \mathcal{R}(r,\theta)~ dS
\end{align}
where $C_2$ is the positive pre-factor
\begin{align}
	C_2 = \frac{\epsilon_o \gamma_e M_a^2 }{ 4 ~ \Omega_a} \cdot A_P(t)  ~.
\end{align}
Substituting the overlap integral $\mathcal{I}_{\textrm{overlap}} > 0$ 
and the relative phase $\psi = \pi/2$ determined from the synchronous solution to the optical wave equation \ref{eqn:rhodot4}, 
we have the simplified form
\begin{align}
	\label{eqn:rhodot3}
	\rho'(t) \approx - \frac{C_2 ~ A_{aS}(t)}{\mathfrak{R}_{opt}^2} \, \mathcal{I}_{\textrm{overlap}}   ~.
\end{align}
which indicates that $\rho'(t)$ is always negative.

A major outcome of this calculation is now revealed. The synchronous solution of the two wave equations \ref{eqn:OpticalWave} and \ref{eqn:AcousticWaveEqn} implies that when both the pump and anti-Stokes optical waves resonate (equations \ref{eqn:TEP}, \ref{eqn:TEaS}), the acoustical resonance (equation \ref{eqn:AcousticWave}) is attenuated.
\\

\subsection{Verification of acoustic gain during Stokes scattering}

We will now confirm that the energy flow in \ref{eqn:rhodot3} is inverted (i.e.\ goes from light to sound) if the Stokes wave is considered instead of the anti-Stokes wave in \ref{eqn:TEaS}. 
We emphasize that in our cooling experiment the Stokes line was off-resonantly attenuated and that this section is only for mathematical confirmation.

A major difference here is that the Stokes line has frequency $\omega_S$ lower than the pump (as opposed to the anti-Stokes line having higher frequency). Following the method in section \ref{sec:Optical} for a synchronous solution to the Stokes wave defined as
\begin{align}
	\tilde{E_S} & = A_S(t) ~ E_S(r,\theta) ~ \cos(M_S ~ \phi - \omega_S ~ t + \psi) 
\end{align}
we obtain the same result for the phase angle $\psi = \pi/2$. At this stage there is no difference between the solutions for Stokes and anti-Stokes scattering.

In the acoustic wave equation solved in section \ref{sec:Acoustic}, however, we obtain a different result. Searching for a synchronous solution in equation \ref{eqn:AcousticRHS} results in a change of sign for the $\psi$ term in the electrostrictive pressure
\begin{align}
\left. \tilde p_{st} \right|_{\textrm{at } \Omega_a}  = - \frac{1}{2} \epsilon_o \gamma_e A_P(t) A_{aS}(t) E_P(r,\theta) E_{aS}(r,\theta) \cos(M_a ~ \phi -\Omega_a ~ t - \psi) ~.
\end{align}
This is a major difference that eventually results in a sign inversion in the rate-of-change of acoustic wave amplitude.
\begin{align}
	\rho'(t) \approx + \frac{C_2 ~ A_S(t)}{\mathfrak{R}_{opt}^2} \, \mathcal{I}_{\textrm{overlap}}   ~.
\end{align}
As expected, this result indicates that acoustic wave sees gain (or, is heated) through Stokes scattering, which is consistent with previous derivations of \citeS{Boyd}, \citeS{Yariv} and \citeS{PhysRevLett.12.592}.
\\

\subsection{Cooled acoustical mode}
\label{sec:CooledMode}

We now incorporate the loss of anti-Stokes photons from the cavity into equation \ref{eqn:Aas_prime_simple}, and we set $A'_{aS}(t) = 0$ for steady state.
\begin{align}
	C_1 ~ \rho(t)  ~ \mathcal{I}_{\textrm{overlap}} - \delta_{aS} A_{aS}(t) = ~ 0
\label{eqn:Aas_ss}
\end{align}
where $\delta_{aS}$ is a loss rate defined by the loaded optical quality factor \citeS{GorodetskyOptimal} for the anti-Stokes optical mode ($\delta_{aS} = \omega_{aS}/2 Q_{aS}$). This provides a solution for the steady state anti-Stokes mode amplitude for a given acoustical density fluctuation.
\begin{align}
	A_{aS}(t) = \frac{C_1 ~ \mathcal{I}_{\textrm{overlap}}}{\delta_{aS}} \rho(t) ~.
	\label{eqn:Aas_ss_soln}
\end{align}
Substituting \ref{eqn:Aas_ss_soln} in \ref{eqn:rhodot3} reveals that
\begin{align}
	\rho'(t) & = - \frac{C_1 \, C_2 \, \mathcal{I}_{\textrm{overlap}}^2}{\mathfrak{R}_{opt}^2 \, \delta_{aS}} \rho(t) \notag \\
	& = - \left(\frac{\gamma_e^2 ~\epsilon_o ~\mathcal{I}_{\textrm{overlap}}^2 ~M_a^2 ~\omega_{aS}^2 }{8 ~c^2  ~\rho_o  ~M_{aS}^2 ~\Omega_a} \cdot Q_{aS} A_p^2 \right) \rho(t) ~ \equiv ~ - \, \frac{\Gamma_\textrm{Brill.}}{2} \, \rho(t)
	\label{eqn:rhodot_final}
\end{align}
where we defined the R.H.S. with $\Gamma_\textrm{Brill.}$ being the rate of Brillouin cooling by light.

We now take into account intrinsic mechanical damping $\Gamma_a$ and thermal
fluctuations that enter the equation of motion for the quadrature amplitude
$\rho$ as a Langevin noise force term \citeS{StatMech_vonKampen}. 
The solution of the equation of motion reveals that the acoustical density-fluctuation $\tilde\rho$ has a power spectral density of Lorentzian shape, with linewidth $\Gamma_a + \Gamma_\textrm{Brill.}$.
As a result, we obtain
\begin{align}
	\frac{T_\textrm{eff}}{T_\textrm{r.t.}} ~ = ~ \frac{\langle \tilde\rho^2 \rangle}{\langle \tilde\rho^2_{\textrm{r.t.}} \rangle} ~ = ~ \frac{\Gamma_a}{\Gamma_a + \Gamma_\textrm{Brill.}}
	\label{eqn:ratios}
\end{align}
where $\langle ~ \rangle$ represents time average, and $\langle \tilde\rho^2_{\textrm{r.t.}} \rangle$ is the Brownian density-fluctuation at room temperature $T_\textrm{r.t.}$.
Here $T_\textrm{eff}$ is the effective temperature of the acoustical mode.
Previous studies \citeS{PhysRevLett.83.3174}, \citeS{Metzger:2004p1357}, \citeS{Steeneken:2011p1358}, \citeS{Arcizet:2006p1092} have also arrived at similar expressions relating linewidth to temperature.
%
% \\

We can substitute the pump field-intensity $A_P$ in the cavity (equation \ref{eqn:rhodot_final}) with input pump power $P_\textrm{input}$ instead, as 
\begin{align}
	P_\textrm{input} = \frac{M_P}{2 \,\eta \,Q_P} A_P^2
\end{align}
where $Q_P$ is the quality-factor of the pump mode, and $\eta$ is the vacuum impedance. We then write equation \ref{eqn:ratios} using \ref{eqn:rhodot_final} as
\begin{align}
	\frac{T_\textrm{eff}}{T_\textrm{r.t.}} ~ = ~ \cfrac{1}{1 ~ + ~ \cfrac{\gamma_e^2 ~\mathcal{I}_{\textrm{overlap}}^2}{2 ~c^3  ~\rho_o} \cdot \cfrac{M_a^2 ~\omega_{aS}^2 }{M_P ~M_{aS}^2 ~\Omega_a^2} \cdot Q_a \, Q_{aS} \, Q_P \, P_\textrm{input}}
\end{align}
where we used the relationship between acoustical-loss and acoustical quality-factor $\Gamma_a = \Omega_a / Q_a$.

This result shows that the cooling due to the anti-Stokes process improves with increasing pump optical power $P_\textrm{input}$. Increasing optical quality factors $Q_{aS}$, $Q_P$, and the acoustical quality factor $Q_a$, also increases the cooling efficiency. 
\\

\section{Experimental measurement of the effective acoustical mode temperature}

We experimentally measure the beat note between the pump optical signal and the anti-Stokes scattered signal on a photodetector. We will now discuss how this beat allows us to observe the acoustical mode and measure the effective temperature $T_\textrm{eff}$.

We first rewrite \ref{eqn:Aas_ss_soln} as
\begin{align}
	A_{aS} = \frac{\mu_o \epsilon_o \gamma_e ~ \omega_{aS}^3 ~ \mathfrak{R}_{opt}^2 ~ \mathcal{I}_{\textrm{overlap}}} { 4 ~ \rho_o ~ M_{aS}^2 ~ \delta_{aS}} \cdot A_P \cdot \rho ~ .
	\label{eqn:RewriteAas}
\end{align}
As a result, the 
acoustical power spectral density (of $\rho$) is also transferred to the optical signal $A_{aS}$ around the anti-Stokes frequency $\omega_{aS}$.
As the anti-Stokes signal now carries the spectral profile of the acoustical mode, its beat note with the pump optical signal also has the same spectral shape.

We then define the reflectance of the pump power $P_P \propto A_P^2$ into the anti-Stokes mode $P_{aS} \propto A_{aS}^2$ as $R$. Using \ref{eqn:RewriteAas} we get 
\begin{align}
	\label{eqn:Rprop}
	R = \frac{P_{aS}}{P_P} \propto \rho^2 
\end{align}
where the calculation is for powers inside the resonator.
The power reflectance $R$ is expected to have a Lorentzian spectrum owing to its proportionality with $\rho^2$. 
Experimentally, we measure the spectrum of $R$ through the beat note between the pump and anti-Stokes signals, and we calculate the linewidth $\Gamma_\textrm{eff} = \Gamma_a + \Gamma_\textrm{Brill.}$ of the cooled density fluctuation. 

We then calculate the effective mode temperature $T_\textrm{eff}$ through equation \ref{eqn:ratios}.
In our experiment, $T_\textrm{r.t.}$ was 294 K, and we measured the room-temperature linewidth of the observed 95 MHz mode to be $\Gamma_a = 7.7$ kHz (see Figure 4b). 
\\

\section{Brownian amplitude of the acoustical mode}
\label{sec:BrownianAmplitude}

The thermal equilibrium occupation of phonons ($n_k$) in an acoustical mode of frequency $\Omega_a$ is provided by the expression (see~\citeS{Kittel} chapter 5) 
\begin{align}
	n_k = \left[ e^{\hbar \Omega_a / k_B T} - 1 \right]^{-1}
\end{align}
where $k_B$ is Boltzmann's constant, and $T$ is the ambient temperature.
Additionally, the amplitude of displacement ($u_o$, as shown in figure \ref{fig:Mode}) from the equilibrium position 
due to the acoustical mode is quantized (see~\citeS{Kittel} chapter 4) based on the expression
\begin{align}
	u_o^2 = 4 \left( n_k + \frac{1}{2} \right) \frac{\hbar}{m_{\textrm{eff}}\, \Omega_a}
\end{align}
where $m_{\textrm{eff}}$ is the effective mass of the mode (calculated numerically). 
Using the thermal equilibrium phonon occupancy at room temperature, the above equations can be simplified to 
\begin{align}
	u_o^2 \approx \frac{4 k_B T}{m_{\textrm{eff}} \, \Omega_a^2 } ~.
\end{align}
The mode amplitude for the $M_a=12$, $\Omega_a = 95$ MHz mode shown in section \ref{sec:Simulation} is calculated to be $u_o = 9.57~\textrm{fm}$ at 300 K.
\\

%%%%%%%%%%%%%%%%%%%%%%%%%%%%%%%%%
%\bibliographystyleS{IEEEtran}
\bibliographystyleS{plain}
\bibliographyS{BrillCool}
%\input{S.bbl}
%%%%%%%%%%%%%%%%%%%%%%%%%%%%%%%%%

\end{document}